\documentclass[twoside]{dis09}
\usepackage[latin1]{inputenc}
\usepackage[dvips]{graphicx,epsfig,color}
\usepackage{wrapfig,rotating}
\usepackage{amssymb,amsmath,array}
\usepackage{bm}

\newcommand{\be}{\begin{equation}}
\newcommand{\ee}{\end{equation}}
\newcommand{\bea}{\begin{eqnarray}}
\newcommand{\eea}{\end{eqnarray}}
\newcommand{\pup}{p^\uparrow}
\newcommand{\qup}{q^\uparrow}

\begin{document}
\title{Single spin asymmetries in $\bm{\ell p\to h + X}$ processes\thanks{Talk delivered by
U.~D'Alesio at the ``XVII International Workshop on Deep-Inelastic
Scattering and Related Subjects" DIS 2009, 26-30 April 2009, Madrid,
Spain.}}

%***********************************************************************
% AUTHORS INFORMATION AREA
%***********************************************************************
\author{
M. Anselmino$^{1,2}$, M. Boglione$^{1,2}$, U. D'Alesio$^{3,4}$\,, S.
Melis$^{2,5}$, F. Murgia$^{4}$, A. Prokudin$^{1,2}$
\vspace{.2cm}\\
%
% Addresses and institutions
1- Universit\`a di Torino - Dipartimento di Fisica Teorica \\
Via P. Giuria 1, I-10125 Torino - Italy
\vspace{.1cm}\\
2- INFN - Sezione di Torino  \\
Via P. Giuria 1, I-10125 Torino - Italy
\vspace{.1cm}\\
3- Universit\`a di Cagliari - Dipartimento di Fisica \\
Cittadella Universitaria, I-09042 Monserrato (CA) - Italy
\vspace{.1cm}\\
4- INFN - Sezione di Cagliari \\
C.P. 170, I-09042 Monserrato (CA) - Italy
\vspace{.1cm}\\
5- Universit\`a del Piemonte Orientale - Dipartimento di Scienze e
Tecnologie Avanzate
\\ Viale T. Michel 11, I-15121 Alessandria -
Italy }
%
%***********************************************************************
% END OF AUTHORS INFORMATION AREA
%***********************************************************************

\maketitle

\begin{abstract}
We study the transverse single spin asymmetry (SSA), $A_N$, for the
single inclusive process $\ell p^\uparrow\to h+X$, in a perturbative
QCD factorization scheme with inclusion of spin and transverse
momentum dependent (TMD) distributions. By adopting the relevant
TMDs (Sivers and Collins functions) as extracted from semi-inclusive
deep inelastic scattering (SIDIS) and $e^+e^-$ data, predictions for
these SSAs are given. A measurement of $A_N$ for this process could
then provide a direct test of the validity of the TMD factorization.
\end{abstract}

\section{Introduction}
In this contribution~\cite{url} we discuss a preliminary study of
transverse SSAs in inclusive hadron production in lepton-proton
collisions as a tool to test the TMD factorization hypothesis.

Azimuthal and transverse single spin asymmetries, with their
behaviour and size, definitely represent a challenge for the QCD
factorization theorems and, at the same time, have opened a new way
to learn on the spin structure of hadrons. A large amount of data
has been analyzed by many experimental collaborations in various
inclusive processes (for reviews see, e.g.,
Ref.~\cite{D'Alesio:2007jt}): the left-right asymmetries in single
polarized proton-proton collisions (from the early E704 to the
latest STAR and BRAHMS data), the azimuthal asymmetries in single
polarized SIDIS (HERMES and COMPASS) and in hadron-pair production
in $e^+e^-$ annihilation (Belle).

Two theoretical approaches have been proposed to describe such
phenomena \cite{D'Alesio:2007jt}: $i)$ a generalization of the pQCD
factorization scheme with inclusion of a new class of spin and TMD
distribution and fragmentation functions; $ii)$ a collinear QCD
factorization theorem in terms of higher-twist parton correlation
functions.

The first approach was initially adopted~\cite{Anselmino:1994tv} to
explain the early SSAs observed in $pp\to \pi +X$ and later applied,
with success, to predict~\cite{D'Alesio:2004up} $A_N$ at much larger
energies. An extended and systematic study can be found in
Ref.~\cite{Anselmino:2005sh}. Presently, the TMD approach is
believed to hold for SSAs characterized by the presence of two
scales, a large (i.e. $Q^2$) and a small one ($p_T\simeq
\Lambda_{\rm QCD}$), like for the production of small $p_T$ hadrons
in SIDIS processes or of small $p_T$ lepton-pairs in Drell-Yan
processes~\cite{Ji:2004wu,Ji:2004xq}. For a detailed and complete
classification of all TMDs and their role in azimuthal and SSAs in
SIDIS see Ref.~\cite{Bacchetta:2006tn}.

Based on these results the first extractions of the Sivers, Collins
and transversity functions have been
performed~\cite{Anselmino:2005ea, Vogelsang:2005cs, Collins:2005ie,
Efremov:2006qm, Anselmino:2007fs, Anselmino:2008sga,
Anselmino:2008jk}. Notice that all these analyses of SIDIS data are
carried out in the $\gamma^* - p$ c.m.~frame, according to the
following TMD factorization formula:
 \be
 d\sigma^{\ell p \to \ell' h+ X} \sim \sum_q f_{q/p}(x, k_\perp;
Q^2) \otimes d\hat\sigma^{\ell q \to \ell q} \otimes D_{h/q}(z,
p_\perp; Q^2)\,,
 \ee
where $\bm{k}_\perp$ and $\bm{p}_\perp$ are, respectively, the
transverse momentum of the quark in the proton and of the final
hadron with respect to the fragmenting quark.

The alternative formalism, the higher-twist approach, has been
proved to hold for SSAs where a single hard scale is relevant, like
the inclusive production of large $p_T$ hadrons in hadron-hadron
collisions~\cite{Qiu:1991pp}. A corresponding and rich phenomenology
has been developed in Refs.~\cite{Qiu:1998ia, Kanazawa:2000kp,
Kouvaris:2006zy}.

In the last years these two approaches have been shown to be
somewhat related and equivalent in the kinematic regime where they
both apply~\cite{Ji:2006ub}.

However, a definite proof of the validity of the TMD factorization
for SSAs in inclusive particle production in hadron-hadron
collisions, with a single large scale, is still lacking. In this
context we mention that a sort of modified TMD factorization
approach has been discussed in the study of dijet production at
large $p_T$ in $pp$ collisions by including the proper gauge-links
structure in the elementary
processes~\cite{Collins:2007nk,Vogelsang:2007jk,Bomhof:2007xt}.

What we aim at discussing here is then an experimental test of the
TMD factorization by considering the SSA in the $\ell p\to h +X$
process, where a single large $p_T$ final particle is
detected~\cite{ Anselmino:2009xx}. This is the analogue of the
left-right asymmetry observed in $pp\to h +X$. Differently from
SIDIS, we will now perform the analysis in the lepton-proton
c.m.~frame without the detection of the final lepton, but still
requiring a large $Q^2$ regime (see below).

Similar studies, although with different motivations, were presented
in Refs.~\cite{Anselmino:1999gd, Koike:2002gm, She:2008tu}.

\section{TMD approach to $\bm{\ell p\to h + X}$}

\begin{wrapfigure}{r}{0.35\columnwidth}
\vspace*{-1.2cm}
\centerline{\includegraphics[width=0.25\columnwidth]{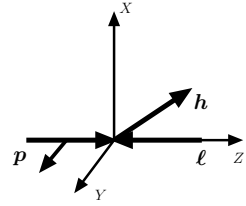}}
\caption{Kinematics.}\label{Fig:kin}
\end{wrapfigure}

We consider the process $p^\uparrow\ell\to h + X$, with $p$ moving
along the positive $Z$-axis in the $p-\ell$ c.m.~frame, the final
hadron produced in the $X-Z$ plane and $\uparrow$ direction along
$+Y$ (Fig.~\ref{Fig:kin}). We assume the same TMD factorization
scheme as for the process $p^\uparrow p\to h +X$ and compute
$A_N=(d\sigma^\uparrow-d\sigma^\downarrow)/(d\sigma^\uparrow+d\sigma^\downarrow)$.
The numerator of this SSA reads
 \bea
 \label{master}
 d\sigma^\uparrow - d\sigma^\downarrow
  &\sim & \sum_{q} \Bigl\{
 \Delta^N\!f_{q/\pup} \cos\phi_q \otimes d\hat\sigma \otimes
 D_{h/q} \nonumber \\
&+& h_1^{q/p} \otimes d\Delta\hat\sigma \otimes \Delta^N\!
D_{h/q^\uparrow}\cos\phi_C  + h_{1T}^{\perp q/p} \otimes
d\Delta\hat\sigma \otimes \Delta^N\! D_{h/q^\uparrow}
\cos(\phi_C-2\phi_q) \Bigr\} \,,\nonumber\\
 \eea
where the $q\ell\to q\ell $ elementary cross sections are given by
 \be
 \label{sigma}
 d\hat\sigma \propto  e_q^2\,\frac{\hat s^2+ \hat u^2}{2\,\hat
t^2} \hspace*{2cm} d\Delta \hat\sigma \propto - e_q^2\, \frac{\hat s
\hat u}{\hat t^2}\,,
 \ee
with $\hat s, \hat t, \hat u$ the Mandelstam invariants and
$\phi_C\equiv \phi_h^H + \phi_{q'}$, with $\phi_q (\phi_{q'})$ the
azimuthal angle of the initial (fragmenting) quark in the c.m.~frame
and $\phi_h^H $ the hadron azimuthal angle in the outgoing-quark
helicity frame. Notice that the result shown in Eq.~(\ref{master})
can be also obtained as a particular case of the $(A,S_A)+(B,S_B)\to
C+X$ process~\cite{Anselmino:2005sh}.
In equation~(\ref{master}) we recognize the contributions from the
Sivers function~\cite{Sivers:1989cc,Sivers:1990fh} (first line)
 \be
\hat f_{q/\pup}(x, \bm{k}_{\perp}) - \hat f_{q/p^\downarrow}(x,
\bm{k}_{\perp}) \equiv \Delta^N\! f_{q/\pup}\,(x, k_{\perp}) \,
\bm{S} \cdot (\hat{\bm{p}} \times \hat{\bm{k}}_{\perp })
\label{siv}\,,
 \ee
where the mixed product in our configuration gives the $\cos\phi_q$
dependence in Eq.~(\ref{master}), and from the Collins
function~\cite{Collins:1992kk} (second line)
 \be
 \hat D_{h/\qup}(z,\bm{p}_\perp) - \hat D_{h/q^\downarrow}(z,\bm{p}_\perp)
 \equiv \Delta^N\! D_{h/q^\uparrow}(z,p_\perp)\, \bm{s}_q \cdot (\hat{\bm{p}}_q \times
 \hat{\bm{p}}_\perp)\,,
 \label{col}
 \ee
coupled with $h_1$ (the TMD transversity distribution) and
$h_{1T}^\perp$; in this case the azimuthal dependences in
Eq.~(\ref{master}) arise from the phases entering the TMD
distributions, the elementary scattering and the Collins function.

Before presenting our results we want to comment on some aspects
peculiar for this process. First of all, as in SIDIS, we have a
single partonic subprocess and only the $\hat t$-channel
contribution (much simpler than the $pp\to h +X$ case). However,
without the detection of the final lepton one is not able to
reconstruct the lepton plane and then to access separately the
Sivers and the Collins effect (as usually done for SIDIS). On the
other hand in the backward region (w.r.t.~the proton direction)
$|\hat u|$ becomes smaller and so does the spin transfer
$d\Delta\hat\sigma/d\hat\sigma$ (see Eq.~(\ref{sigma})) implying a
strong dynamical suppression of the Collins effect. At the same
time, this does not affect the Sivers contribution since, contrary
to what happens in $pp\to h +X$, no $\hat u$-channel is active .
[Remember also that the $\hat t$ variable strongly depends on
$\phi_q$ (the azimuthal dependence of the Sivers effect, see
Eq.~(\ref{master}))].

It is also well known that the $\ell p\to h +X$ process is dominated
by quasi-real photon-exchange with low $p_T$ hadrons in the final
state. However, for the validity of the adopted perturbative QCD
factorization scheme the elementary scattering must be governed by a
large momentum transfer, let us say $Q^2>1$ GeV$^2$. This is
trivially guaranteed in the usual collinear configuration by
requiring a moderate-large $p_T$ for the final hadron ($p_T\ge 1$
GeV). We have checked that, by including $k_\perp$ effects in the
kinematics, the safe region ($Q^2> 1$ GeV$^2$) corresponds to a
minimum $p_T$ value of the order of 1.5 GeV. More precisely, for
such value one has to restrict to the backward region
($x_F=2p_L/\sqrt s<0$ ), while, at larger values, like $p_T=2.5$
GeV, we can explore the full $x_F$ range, remaining in the large
$Q^2$ regime. These are the values considered in our estimates.

A word of caution has still to be added. In our calculation we do
not include any hard gluon emission which could give a large $p_T$
hadron even in the low $Q^2$ regime. This
 would lead to a two-jet
event and might be experimentally excluded by requiring the absence
of any hadron activity in the opposite hemisphere w.r.t.~the
detected hadron.

\section{Results}
We show our estimates for two different kinematic setups related
to the ongoing HERMES ($p_{\rm Lab}= 27.5$ GeV) and COMPASS ($p_{\rm
Lab}= 160$ GeV) experiments and focus on pion production. By
numerical calculation we have checked that the (maximized)
contribution to $A_N$ coming from $h_{1T}^\perp$ (see
Eq.~(\ref{master})) is totally negligible, as well as the analogous
contribution to the unpolarized cross section involving the
Boer-Mulders function. In our estimates we will adopt the present
information on the relevant TMDs (Sivers, Collins and transversity
functions) as extracted in the latest analyses of SIDIS and $e^+e^-$
data~\cite{Anselmino:2008sga, Anselmino:2008jk}. We will show
separately the contribution to $A_N$ from the Sivers or the Collins
effect alone.

\begin{figure}
\centerline{
\includegraphics[width=0.26\columnwidth,angle=-90]{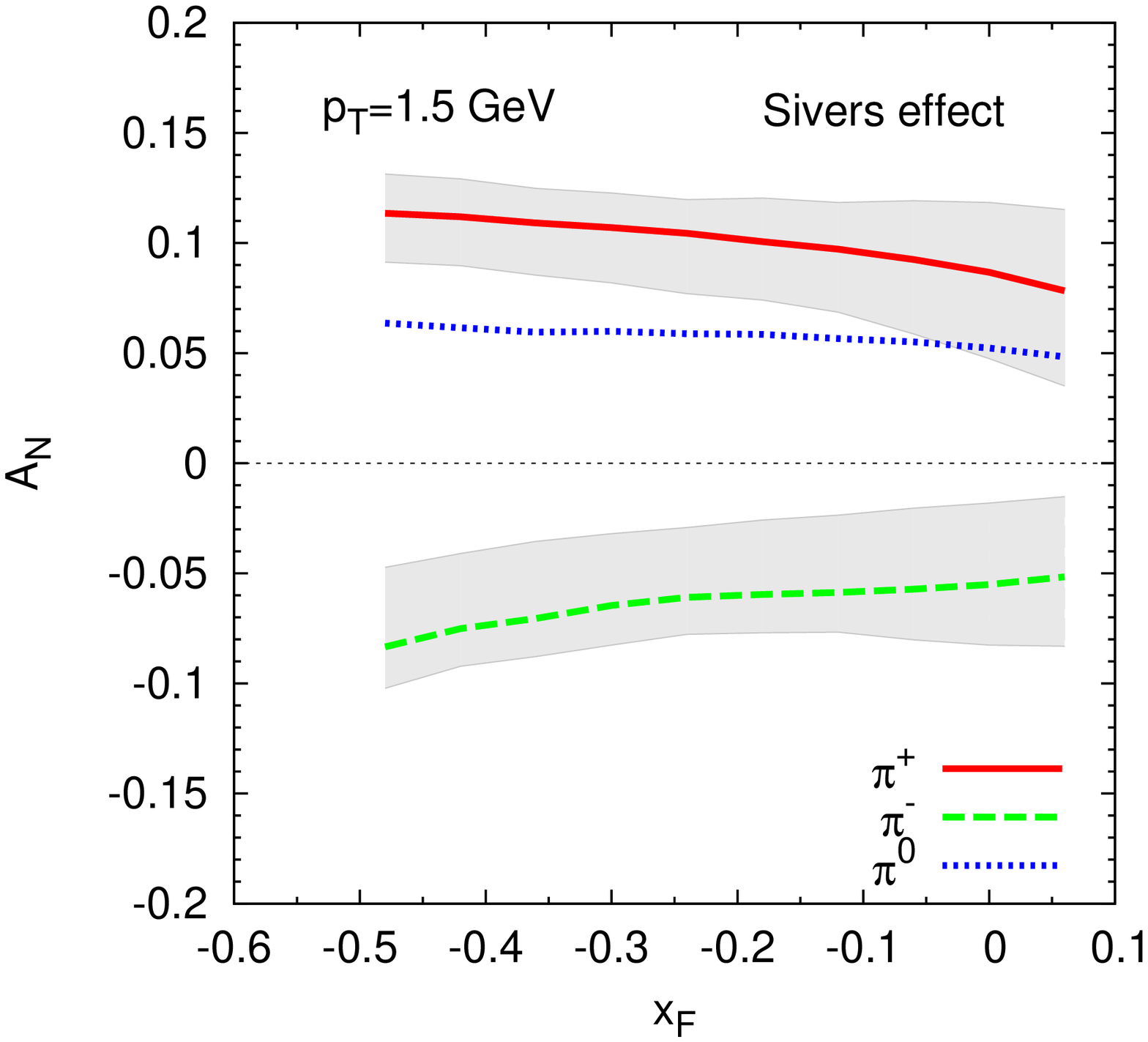}
\includegraphics[width=0.25\columnwidth,angle=-90]{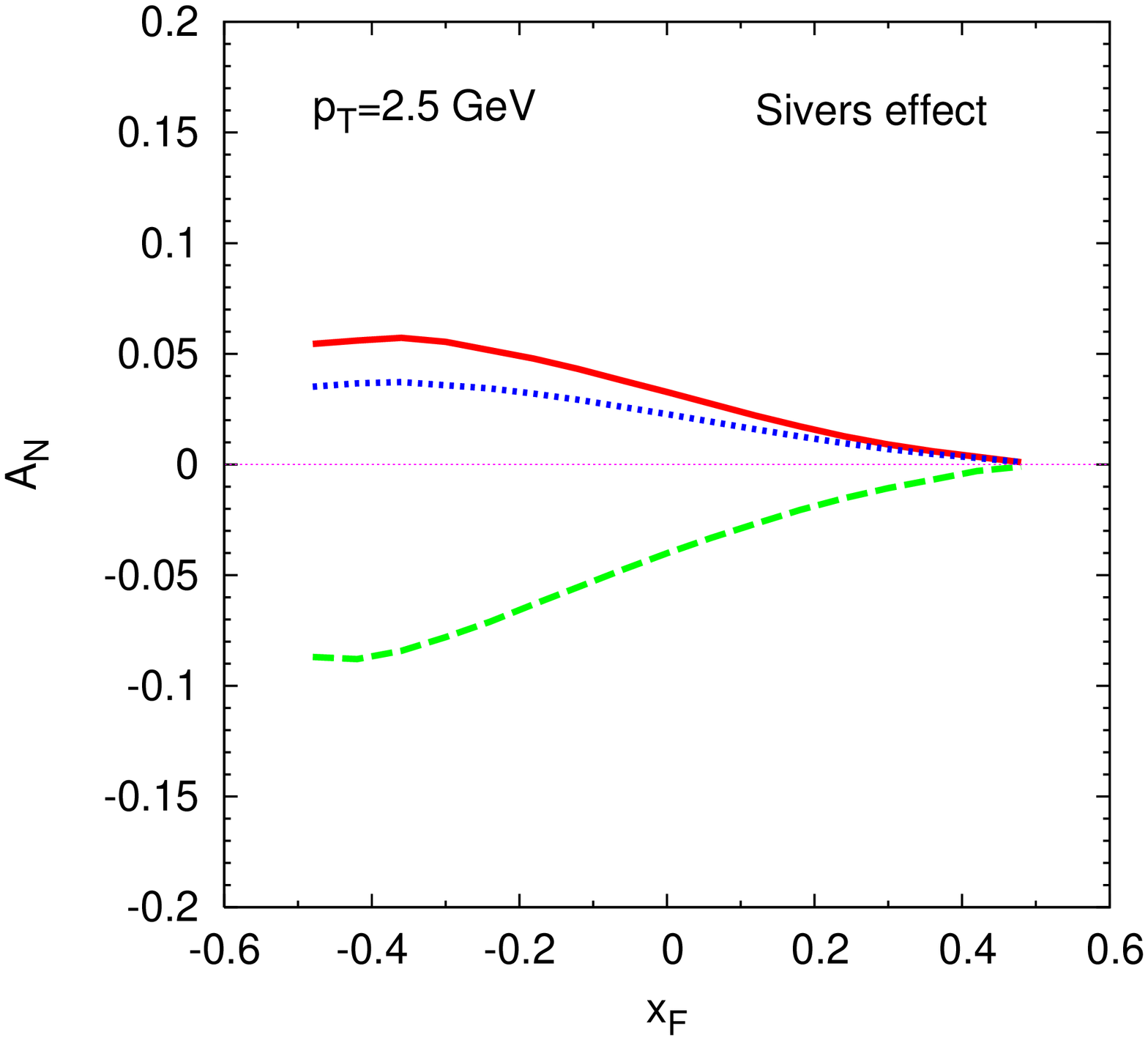}
\includegraphics[width=0.25\columnwidth,angle=-90]{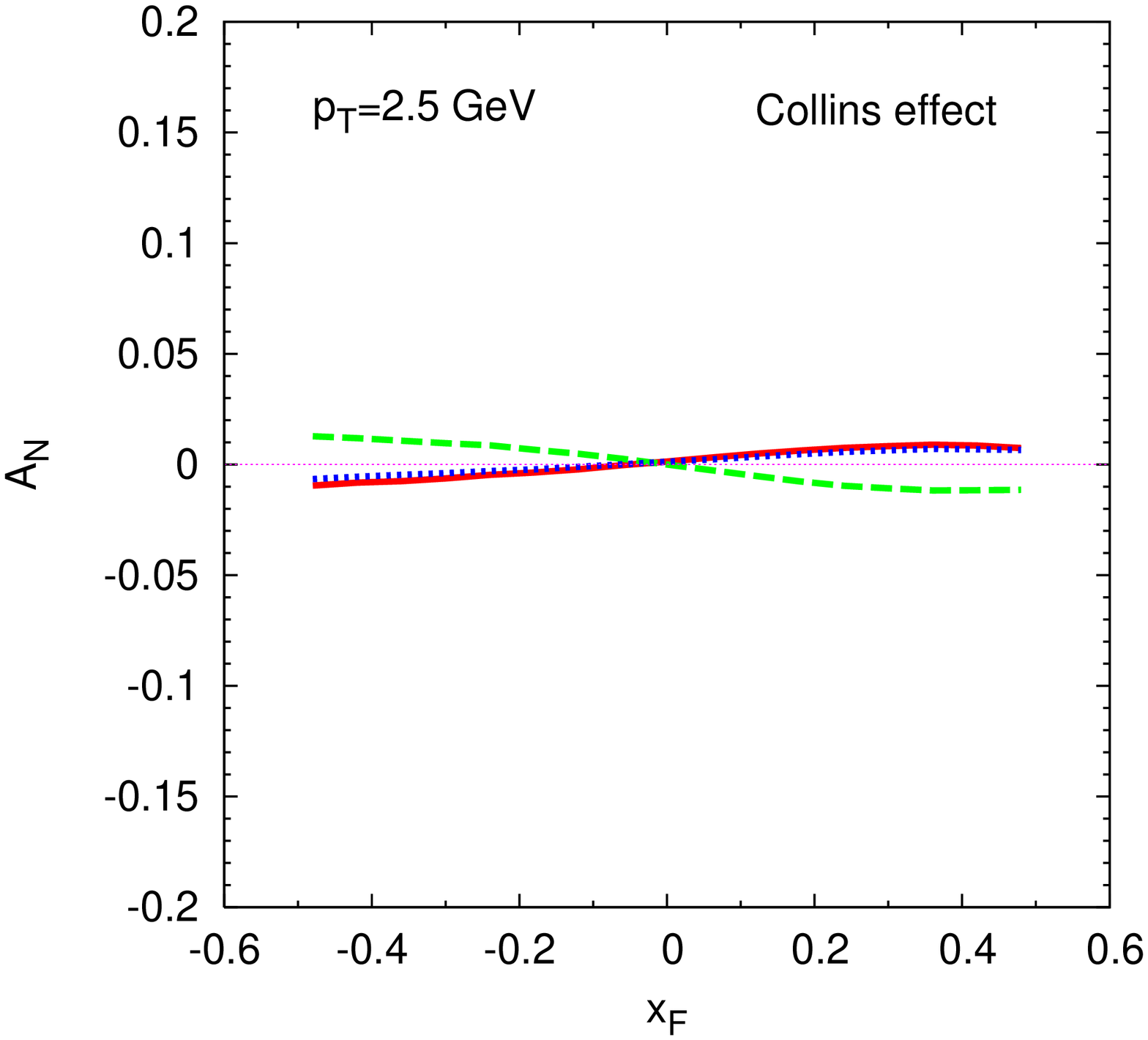}}
\caption{$A_N(p^\uparrow\ell \to \pi +X)$ for HERMES kinematics
($p_{\rm Lab}= 27.5$ GeV) at fixed $p_T$ values as a function of
$x_F$. Left and central panels: Sivers effect (the statistical
uncertainty bands for charged pions at $p_T=1.5$ GeV are also
shown); right panel: Collins effect.} \label{Fig:hermes}
\end{figure}

In Figure~\ref{Fig:hermes} (left and central panels) we present our
estimates for the Sivers contribution to $A_N$ at HERMES kinematics
for two values of $p_T$. As discussed above, for $p_T=1.5$ GeV only
the backward region can be considered. For charged pion production
at $p_T=1.5$ GeV we also show the statistical uncertainty bands from
the fits. The largest $A_N$ values obtained correspond to the $x$
region (in the polarized proton) where the Sivers functions, for up
and down quarks, reach their maxima. Notice that while for positive
$x_F$, the minimum of $x$ (in the polarized proton) is given,
roughly, by $x_F$, for negative $x_F$ the minimum of $x$ is
controlled by the ratio $p_T/\sqrt s$, with $z$ always bigger than
$|x_F|$. It is interesting to note the sizable $A_N$ for $\pi^-$
production (larger than the corresponding $A_{UT}$ in SIDIS) due to
the dominance of the down-quark contribution with a small
contamination from the up quark.

For $p_T=1.5$ GeV the Collins effect (not shown), involving $h_1$,
is negligible, while it reaches at most 1-2\% at the largest $|x_F|$
values for $p_T=2.5$ GeV (Fig.~\ref{Fig:hermes}, right panel).

\begin{figure}
\centerline{
\includegraphics[width=0.26\columnwidth,angle=-90]{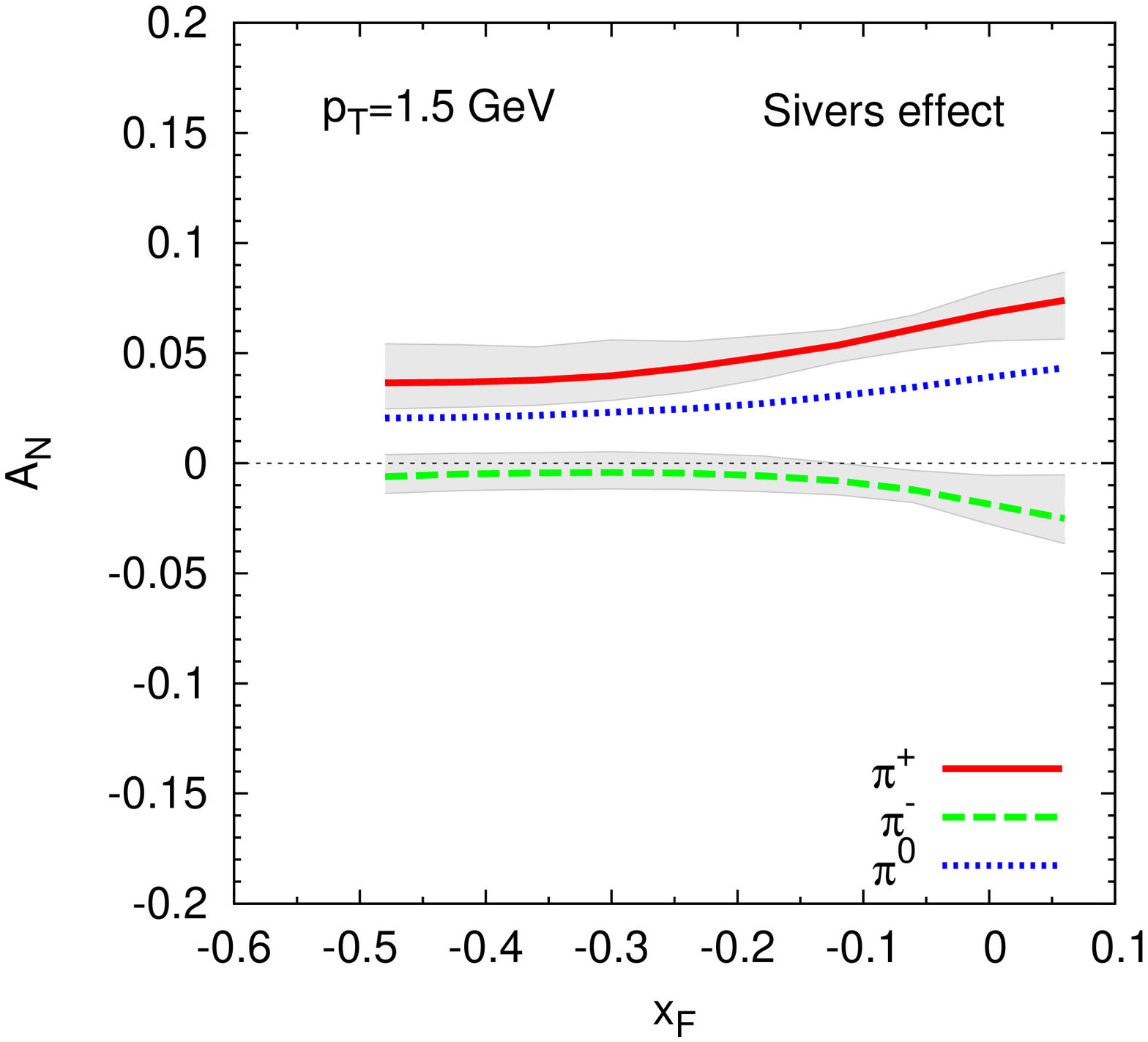}
\includegraphics[width=0.25\columnwidth,angle=-90]{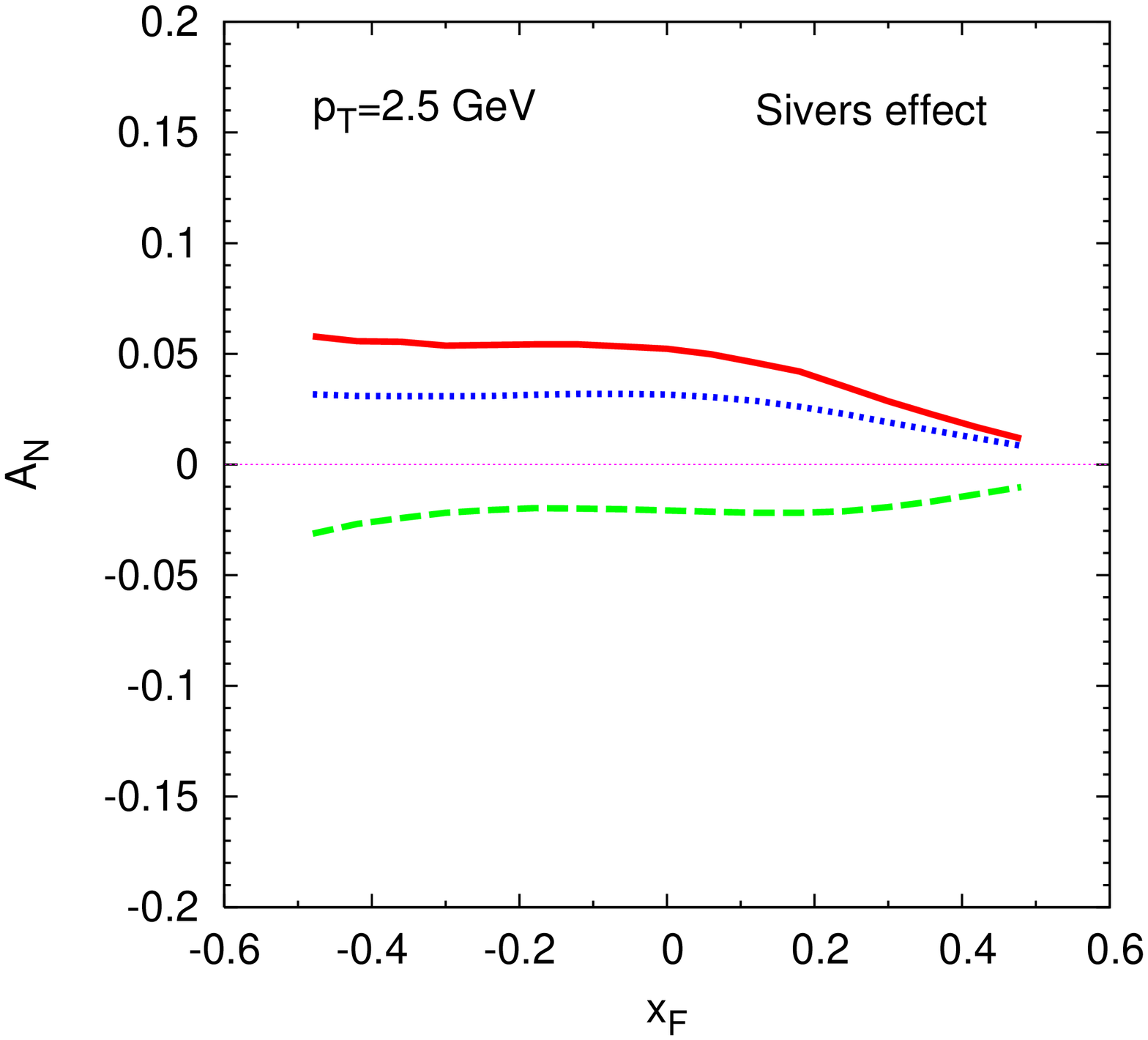}
\includegraphics[width=0.25\columnwidth,angle=-90]{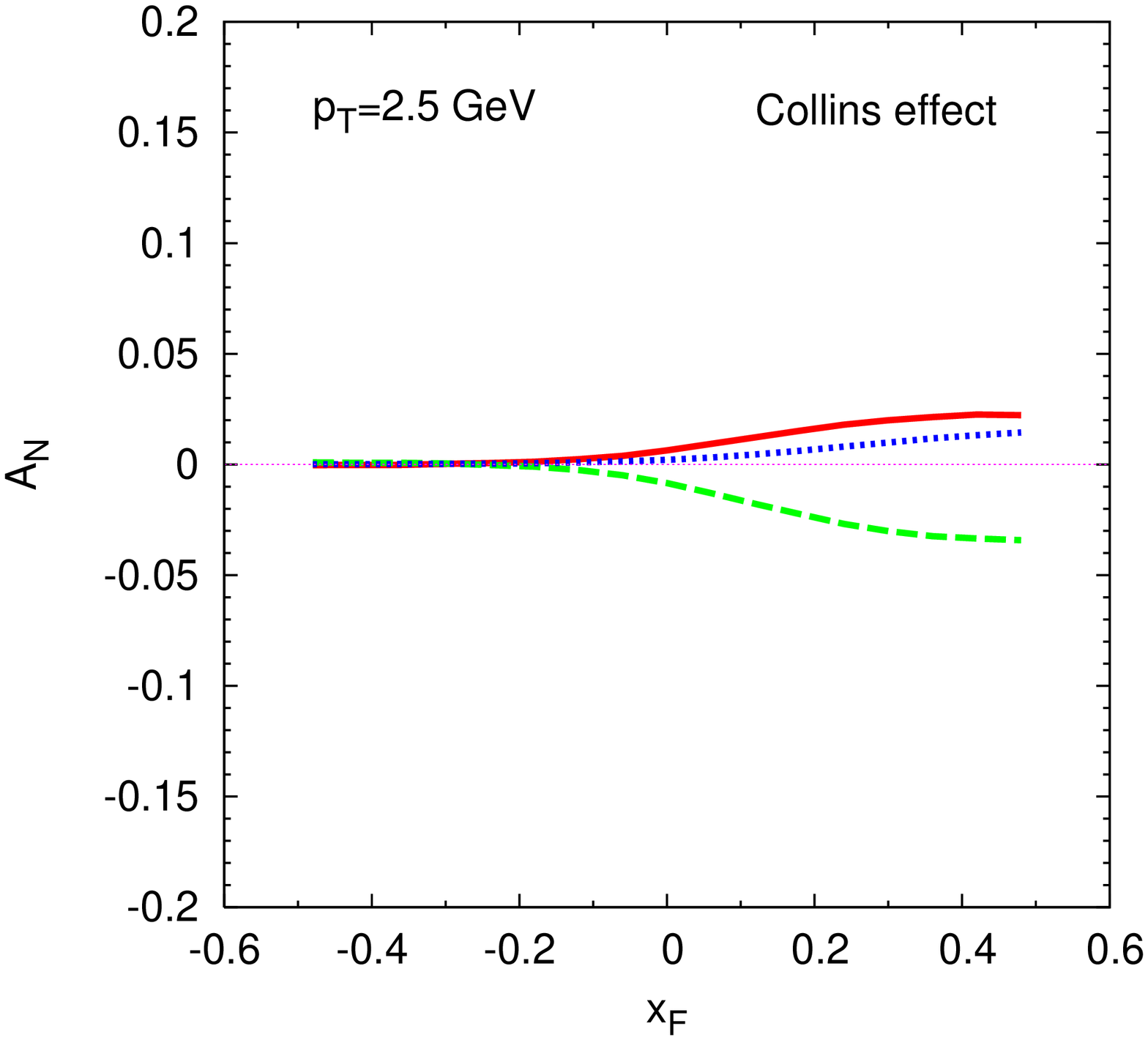}}
\caption{Same as in Fig.~\ref{Fig:hermes} but for COMPASS kinematics
($p_{\rm Lab}= 160$ GeV).
}\label{Fig:compass}
\end{figure}

In Figure~\ref{Fig:compass} we show the analogous results for
COMPASS kinematics. Again at $p_T=1.5$ GeV only the Sivers effect
gives a sizable contribution (left panel), while the Collins effect
(not shown) is compatible with zero. At $p_T=2.5$ GeV the Sivers
effect (central panel) dominates only in the backward region, while
in the forward region the Collins effect (right panel) becomes
sizable.

Notice, however, that for $x_F>0.3$ we are probing the Sivers and
the transversity functions in a region ($x>0.3$) where they are not
constrained by present SIDIS data.

For larger energy values, e.g.~$\sqrt s = 100$ GeV, like those reachable at the
proposed EIC experiments and at the same $p_T$ values as above, we
obtain negligible contributions to the SSAs from both the Sivers and
the Collins effect. This is due to the low $x$ region explored and
the corresponding sea-quark dominance.

\section*{Conclusions}
We have presented a preliminary phenomenological study of SSAs in
$\ell p\to h +X$ within a TMD factorization scheme. This process
presents interesting aspects somewhat in between the SIDIS case (for
which the factorization is generally accepted) and the $pp\to h +X$
case (where it is assumed). In this sense it could represent a clear
test of the TMD factorization hypothesis. By adopting the
parameterizations so far extracted from SIDIS data, sizable SSAs
($\simeq 5-10\%$) can be obtained, mainly due to the Sivers effect
and, to a lesser extent, to the Collins effect.

% ****************************************************************************
% BIBLIOGRAPHY AREA
% ****************************************************************************

\begin{footnotesize}

%\bibliographystyle{unsrt}
%\bibliography{spires-fu}

% }

\end{footnotesize}

% ****************************************************************************
% END OF BIBLIOGRAPHY AREA
% ****************************************************************************

\end{document}